# Variational Autoencoder Analysis of Ising Model Statistical Distributions and Phase Transitions

David Yevick
Department of Physics
University of Waterloo
Waterloo, ON N2L 3G7

**Abstract:** Variational autoencoders employ an encoding neural network to generate a probabilistic representation of a data set within a low-dimensional space of latent variables followed by a decoding stage that maps the latent variables back to the original variable space. Once trained, a statistical ensemble of simulated data realizations can be obtained by randomly assigning values to the latent variables that are subsequently processed by the decoding section of the network. To determine the accuracy of such a procedure when applied to lattice models, an autoencoder is here trained on a thermal equilibrium distribution of Ising spin realizations. When the output of the decoder for synthetic data is interpreted probabilistically, spin realizations can be generated by randomly assigning spin values according to the computed likelihood. The resulting state distribution in energy-magnetization space then qualitatively resembles that of the training samples. However, because correlations between spins are suppressed, the computed energies are unphysically large for low-dimensional latent variable spaces. The features of the learned distributions as a function of temperature, however, provide a qualitative indication of the presence of a phase transition and the distribution of realizations with characteristic cluster sizes.

**Introduction:** Machine learning can simulate the behavior of many physics and engineering systems using standardized and simply adapted program libraries. However, in contrast to standard numerical or analytic methods, which can attain arbitrary accuracy if the system properties and initial conditions are known, machine learning extracts features of a data set through multivariable optimization. Errors therefore arise from both the limited number of data records and the dependence of the optimization procedure on computational metaparameters such as the number and connectivity of the network computing elements. [1][2][3] On the other hand, if the properties of a physical system are partially indeterminate because of intrinsic stochasticity or measurement inaccuracy, machine learning can yield predictions that are at least as accurate as those of deterministic procedures.

Initial applications of machine learning in condensed matter physics employed principal component analysis [4][5][6][7] and neural networks. [8][9][10][11][12][13] These were followed by increasingly sophisticated applications involving both supervised learning from sets of input data and output labels [14][15][16][17][18][19][20] and unsupervised learning in which the network is instead trained on unlabeled data [5,21–29]. Representative supervised learning physics examples include quantum system [30–35] and impurity [36] simulations, predictions of crystal structures [37,38], and density functional approximations. [39] Unsupervised learning procedures include cluster analysis which classifies data according to perceived similarities, and feature extraction that projects the data onto a low-dimensional space while retaining its distinguishing properties. These have been successfully applied, often together with additional information such as locality or symmetry, to identify manifest and hidden order parameters and different phases or states. [5,25,40–44] Additionally, algebraic expressions have been obtained for the symmetries or governing equations associated with a physical system [45][46,47] by algebraically parametrizing the functional form of the single neuron output layer of a Siamese neural network [39] or by similarly parametrizing the highest-order principal component of the relevant data records. [48–52] While such procedures in principle enable the identification of novel system properties, they appear difficult to employ if the desired result cannot be expressed as simple combinations of low order polynomials in the system variables.

This article examines in detail the applicability of variational autoencoders (VAEs) to the Ising model in the vicinity of the phase transition temperature. VAEs are simply implemented with high level machine learning packages, and therefore can be readily adapted to physics problems [53–56][19,57,58], Conversely, VAE methods have been analyzed and enhanced with statistical mechanics algorithms. [59,60]. The VAE is a generative modeling procedure that employs neural networks to encode and then decode the information of the input data set. [61,62] While a VAE employs latent variables similarly to, for example, a restricted Boltzmann machine that employs simple expectation maximization applied to a single hidden layer, the VAE transformations between the physical variables and a restricted set of latent variables employ a multilayer neural network, differential backpropagation and a novel cost function. In particular, the VAE input data is mapped to a bottleneck with a small number of nodes (generally 2) through an encoder typically consisting of one or more neural layers with decreasing node numbers. The output of the encoder, which is termed the latent value representation, is then passed through a decoder network with an expanding number of nodes in each layer until the original number of nodes is attained. These neural networks can consist of either standard dense layers or convolutional layers that preserve local features in the data. The VAE is trained by comparing each output record to the original input. Subsequently, simulated data can be generated by providing appropriately distributed random values for the latent variables and evaluating the corresponding output variables.

A single layer VAE with a linear activation function is analogous to principal component analysis (PCA), which identifies orthogonal linear projections of the data that maximize the variance of each projection. Reconstructing the original data from these projections minimizes the mean squared error such that Gaussian distributed data yields the smallest residual error. Although easily coded, the PCA requires the eigenvalues and eigenvectors of a matrix formed from the data records as is therefore computationally intensive for large input data sets. This article therefore quantifies the accuracy of VAEs trained on the standard two-dimensional Ising model and investigates the properties of the outputs associated with different regions of the latent variables for a range of temperatures around the phase transition temperature.

**Variational Autoencoders:** As indicated above, the initial stage of a variational autoencoder inputs a set of $N$ data records, denoted here by $\mathbf{X}$ into a series of network layers with a diminishing number of neurons termed the encoder. Each data record in the set can be described by a point $\vec{x}$ in a high-dimensional space with coordinates corresponding to, for example, each pixel in an image or each spin in a lattice. The encoder network with parameters $\{\lambda_e\}$ maps each $\vec{x}$ in $\mathbf{X}$ to a point $\vec{z}$ in a lower dimensional space of latent variables. This step is performed in a stochastic manner described by a conditional probability distribution, $E_{\{\lambda_e\}}(\vec{z}|\vec{x})$. A decoder neural net with an increasing number of neurons in successive layers then maps the latent parameters back to a new point $\vec{x}'$ in the original variable space in a manner similarly represented by $D_{\{\lambda_d\}}(\vec{x}'|\vec{z})$. The parameters $\{\lambda_e\}$ and $\{\lambda_d\}$ of the encoder and decoder are trained by minimizing a loss function, $L(\lambda_e, \lambda_d)$ through gradient descent backpropagation. This function incorporates both the difference between the input data and the output of the encoder followed by the decoder and the deviation of the encoder and decoder functions from compact probability distributions in $\vec{z}$. Closely separated points in latent variable space then yield decoder outputs with nearly identical features.

To formulate the first objective of the loss function, the distribution of the data records in $\mathbf{X}$ is associated with a probability distribution function $P(\vec{x})$ termed the evidence. The associated information in nats is given by $-\sum_{m=1}^{N}\log P(\vec{x}_m)$, here simply abbreviated as $-\log P(\vec{x})$. Since an arbitrary point $\vec{z}'$ in the latent space will in general not map to a reconstructed point $\vec{x}'$ in the high-dimensional output space for which $P(\vec{x}')$ is large, the loss function should first include the information in the output image corresponding to

$$L^{(1)}(\lambda,\lambda') = -\mathbb{E}_{z\in E_{\{\lambda_e\}}(\vec{z}|\vec{x})}[\log P(\vec{x}')] \tag{1}$$

In Eq.$(1)$ the subscripted expectation symbol refers to an average over the latent data point distribution generated from the input data records according to the encoder probability distribution function. Accordingly, Eq.$(1)$ quantifies the reconstruction fidelity in reproducing the data records in $\mathbf{X}$ from their truncated representation in the latent space. The VAE accomplishes this in part by training the decoder on latent space representations of the actual data (e.g. the subscript in Eq. $(1)$) which yield large values of $P(\vec{x}')$ rather than decoding random latent space points which would overwhelmingly generate decoder outputs with negligible $P(\vec{x}')$.

The loss function further biases the latent space points that generate the reconstructed data toward a compact posterior probability distribution, $P(\vec{z}) = D_{\{\lambda_d\}}(\vec{z}|\vec{x}')P(\vec{x}')$. Adjacent regions in latent space then yield reconstructed records with slightly different properties, such that if two physical features are associated with two closely separated points in the latent space, positions in the latent space between these points yield synthetic output data that interpolates between the features. Since the encoder is described by a conditional probability distribution, $E_{\{\lambda_e\}}(\vec{z}|\vec{x})$ that will be associated later with a Gaussian distribution, this is accomplished by adding a second contribution, $L^{(2)}(\lambda,\lambda')$, termed the Kullback-Leibler (KL) divergence, between and $E_{\{\lambda_e\}}(\vec{z}|\vec{x})$ and $D_{\{\lambda_d\}}(\vec{z}|\vec{x}')$ to the loss function with

$$\begin{aligned} L^{(2)}(\lambda,\lambda') &= \mathbb{E}_{z\in E_{\{\lambda_e\}}(\vec{z}|\vec{x})}\left[\log E_{\{\lambda_e\}}(\vec{z}|\vec{x}) - \log D_{\{\lambda_d\}}(\vec{z}|\vec{x}')\right] \\ &\equiv \mathbb{E}_{z\in E_{\{\lambda_e\}}(\vec{z}|\vec{x})}\left[\mathbb{KL}\left(E_{\{\lambda_e\}}(\vec{z}|\vec{x}) \parallel \log D_{\{\lambda_d\}}(\vec{z}|\vec{x}')\right)\right] \end{aligned} \tag{2}$$

Eq.$(2)$ expresses in nats the information lost when $E_{\{\lambda_e\}}(\vec{z}|\vec{x})$ is represented by $D_{\{\lambda_d\}}(\vec{z}|\vec{x}')$ and therefore the information distance between the two distributions. The Kullback-Leibler divergence is of the form $-\sum_m p_m \log q_m/p_m$ and is therefore greater than $-\sum_m p_m(q_m/p_m - 1)$ which in turn is positive unless $q$ and $p$ are identical. Accordingly, $L^{(2)}$ is minimized when the decoder inverts the action of the encoder.

To minimize the total loss, $L(\lambda,\lambda') = L^{(1)}(\lambda,\lambda') + L^{(2)}(\lambda,\lambda')$ numerically, the KL divergence is first rewritten in terms of simply evaluated quantities. Applying Bayes rule in the form $P(z)D_{\{\lambda_d\}}(\vec{z}|\vec{x}') = D_{\{\lambda_d\}}(\vec{z}|\vec{x}')P(\vec{x}')$ yields

$$\log P(z) + \log D_{\{\lambda_d\}}(\vec{x}'|\vec{z}) = \log P(x') + \log D_{\{\lambda_d\}}(\vec{z}|\vec{x}') \tag{3}$$

and hence

$$L(\lambda,\lambda') = \mathbb{E}_{z\in E_{\{\lambda_e\}}(\vec{z}|\vec{x})}\left[\mathbb{KL}\left(E_{\{\lambda_e\}}(\vec{z}|\vec{x}) \parallel P(\vec{z})\right)\right] - \mathbb{E}_{z\in E_{\{\lambda_e\}}(\vec{z}|\vec{x})}\left[\log E_{\{\lambda_e\}}(\vec{z}|\vec{x})\right] \tag{4}$$

In Eq.(4) the structure of the VAE in which the encoder maps the data in $\vec{x}$ to the latent variable space $\vec{z}$ and the decoder transforms $\vec{z}$ back to a physical space point $\vec{x}'$ is particularly evident. A variational encoder typically approximates the posterior of the encoder, $E_{\{\lambda_e\}}(\vec{z}|\vec{x}_i)$, for a given data point $\vec{x}_i$ by a point chosen from a Gaussian probability distribution with a mean value given by $\vec{\mu}(x_i,\lambda_e)$ and a variance equal to $\vec{\sigma}(x_i,\lambda_e)$. Similarly, $P(\vec{z})$, the sampling distribution in the space of latent variables of the decoder, is generally equated to a Gaussian function with zero mean and unit variance. As a consequence, input records with similar properties yield localized distributions of latent space values with variances smaller than or of the order of unity. If $E_{\{\lambda_e\}}(\vec{z}|\vec{x}_i)$ and $P(\vec{z})$ are both Gaussian, an analytic evaluation of the KL divergence in Eq.(4) for a $k$ dimensional latent vector space yields

$$\mathbb{E}_{z\in E_{\{\lambda_e\}}(\vec{z}|\vec{x})}\left[\mathbb{KL}\left(E_{\{\lambda_e\}}(\vec{z}|\vec{x}) \parallel P(\vec{z})\right)\right] = -\frac{1}{2}\sum_{i=1}^{k}\left(1+\log\sigma_i^2 - \sigma_i^2 - \mu_i^2\right) \tag{5}$$

in which $\mu_i$ and $\sigma_i$ are the mean and standard deviation, respectively, of the $i$:th component of the coding (e.g. the $i$: th latent variable component). To incorporate neural network-based encoders and decoders that support backpropagation into the above formalism the last layer of the encoder, which outputs latent space variables, is duplicated with the outputs of one of the two branches associated with either $\sigma_i$ or $\log\sigma_i^2$ and that of the second branch with $\mu_i$ for each data record. The KL component of the loss function, Eq.(4), is evaluated from these outputs and combined with the standard neural network reconstruction loss. Further, to generate a normalized Gaussian probability distribution for the latent variable coding of the data, the outputs from the two encoder branches is combined with the output of a normalized Gaussian random number generator according to $\vec{z} \to \vec{z} + \tilde{\vec{z}}$ with

$$\tilde{z}_i = \mu_i + \sigma_i * \mathrm{Normal}(\sigma = 1, \mu = 0) \tag{6}$$

before the decoder is applied. In the VAE, all latent variables are local by construction; that is, the latent variables corresponding to one input realization are independent of other realizations. Therefore, the loss averaged over the input data can be expressed as a sum of the losses over each realization in the data set. Accordingly, stochastic gradient descent can be applied to the parameters $\lambda_e, \lambda_d, \vec{\sigma}, \vec{\mu}$ which can be adjusted for each data point or minibatch or can instead be shared across all datapoints.

Once the VAE is trained, a synthetic output realization can be generated simply by selecting any meaningful set of values for the decoder inputs $z_i$ and calculating the resulting output distribution. If the VAE is trained on sets of differing discrete images, points in latent vector space will generate images that interpolate between the elements of the sets. In standard machine learning applications, this property of the VAE can be employed, for example, to generate facial images that are not associated with a particular person.

**Computational Results:** This paper considers the classical two-dimensional ferromagnetic Ising model on a periodic lattice of $8 \times 8$ spins with unit spins described here by the Hamiltonian

$$H = \frac{J}{2}\sum_{i,j\in\text{nearest neighbors}} S_i S_j \tag{7}$$

with spin variables $S_i = \pm 1$. This enables numerous calculations to be performed with limited computational resources and therefore has been consistently employed as a standard benchmark for the application of neural network procedures to spin systems. [48,63,64] In the remainder of this paper, the temperature and energy will be expressed in normalized units $T/k_b$ and $2E/J$ so that the value of $J$ does not enter the numerical results. Further, energies are referred to the energy of the antiferromagnetic state as spin states at higher temperatures sample both the ferromagnetic and antiferromagnetic regions. For an infinite system, the $Z_2$ inversion symmetry is broken below $T = 2.269$, where the system transitions from a disordered to an ordered phase.

This paper examines two implementations of the VAE. While the synthetic output distributions generated by the decoder for different values of the latent variables is highly dependent on the layer architecture as will be evident from the results of this paper, their general characteristics are effectively identical. The models however employ a relatively small number of network parameters to avoid overfitting, which can lead to divergences or unphysical asymmetries in the energy-magnetization space distributions of the synthetic states. At each temperature, the VAE was trained with 400,000 Ising model spin configurations generated with the Wolff cluster reversal procedure. [65] All VAE layers employed rectified linear (relu) activation functions except for the latent variable layers, which instead employ linear activation functions and the final sigmoid layer. Additional calculations indicated that replacing the rectified linear units with other standard activation functions did not significantly affect the results.

In the dense layer (DL) implementation the $8 \times 8$ input spin values are converted from $\pm 1$ to $0, 1$ and after flattening into a $64 \times 1$ vector are input into a dense 32 neuron layer followed by a dense 8 neuron layer and finally a 16 neuron layer which was found to enhance the quality of the results. The subsequent latent variable layer possesses the standard number, $N_{\text{latent}} = 2$ of neurons unless otherwise stated. Finally, the matched decoder consists of successive dense 16, 8 and 32 neuron layers. To interpret the decoder output as a probability, the final 64 neuron layer employs a sigmoid activation function to output a number between 0 and 1 for each array element which is finally reshaped into a $8 \times 8$ two-dimensional array.

The convolutional layer (CL) procedure instead inserts the $8 \times 8$ array of input values into a single two-dimensional convolutional layer with sixteen $3 \times 3$ filters that employs a stride of two and "same" zero padding resulting in 16 filtered $4 \times 4$ outputs. These are subsequently flattened into a single vector that is coupled to the latent layer. The decoder employs a 16 neuron dense layer to generate 16 values from the latent layer output that are afterwards reshaped into a $4 \times 4$ two dimensional array. A transposed two-dimensional convolutional layer with 16 filters. 'same' padding and a stride of 2 then inverts the corresponding encoder layer and is followed by a $8 \times 8$ single filter two-dimensional convolution layer with sigmoid activation functions.

Two procedures were employed to convert the continuous output of the final sigmoid activation layer into a discrete spin distribution. In the first of these, a positive $+1$ spin was assigned to each array element (lattice site) with a value above 0.5 and the remaining elements were set to $-1$. The second technique generates a random number at each site and assigns the $+1$ spin value to the sites for which the value output by the neural network at the site is larger than the associated random value.

To examine the energy-magnetization distribution of the VAE synthetic realizations, a CL VAE was first trained for 130 training epochs on the Wolff realizations at $T = 3.5$ for which the spin configurations sample both the disordered and ordered phases to an approximately equal extent. As discussed in detail in [66][67][68][69], not only is the distribution at this temperature particularly sensitive to numerical modeling error, but its qualitative features for small spin systems are effectively identical to those exhibited by larger spin systems at corresponding temperatures. Accordingly, 200,000 spin configurations were generated with the trained VAE by assigning random values from a unit normal distribution to each coordinate in the two latent parameter space. The joint histograms constructed from the energies and magnetizations of the spin states generated from the associated decoder outputs when positive and negative spins are assigned to the output neurons with value greater and less than 0.5 respectively (the first procedure of the previous paragraph) are given by the thick lines in Figure 1. The thin lines in the figure represent the corresponding distribution associated with the input states. While the energies of the synthetic states are clearly lower than those of the input realizations, the magnetization distribution is reasonably well reconstructed. Assigning instead probabilities given by the VAE output variables according to the second procedure yields the diagram on the left side of Figure 2. Here the input distribution is then qualitatively reproduced but the state energies are displaced toward larger values. While synthetic output distributions generated from the VAE have been previously displayed, the essential information associated with the energy-magnetization diagram was not presented. [70,71]

Similar behavior has been observed in both the PCA for [48] and in restricted Boltzmann machines (RBM) [64] for limited numbers of principle components and hidden neurons, respectively. Indeed, since the number of states increases with energy, the probability of generating a state of higher energy when the local correlations of the spin system are inaccurately modeled exceeds that of generating a lower energy state. Increasing the number of hidden-layer neurons in the RBM or the number of high-order and hence small spatial frequency PCA components in the PCA markedly improved the reconstruction accuracy and therefore the energy-magnetization distribution of the synthetically generated system realizations in the above references. Similarly, the spatial resolution of the reconstructed states in the VAE improves monotonically with the VAE latent space dimension as evident from the middle and rightmost diagrams in Figure 2, which employ 3 and 10 dimensional latent spaces, respectively.

As the VAE maps input distributions with similar structures to closely separated regions in the latent vector space, synthetic system realizations constructed from a grid of points covering the latent vector space display a series of continuously changing configurations with evolving physical properties. This set of realizations are analogous to a basis into which the input realizations are decomposed. Consequently, information regarding the nature of a phase transition can be inferred from the physical properties of these synthetic states. The four diagrams in Figure 3, for example, display an $10 \times 10$ array where the $(n, m)$:th array element consists of the $8 \times 8$ output probability distributions generated with the convolutional layer (CL) decoder applied to latent coordinates for which the cumulative probability function associated with the normal Gaussian distribution, $\mathbf{Normal}(\boldsymbol{\sigma} = \mathbf{1}, \boldsymbol{\mu} = \mathbf{0})$, equals $0.05 + 0.1n$ and $0.05 + 0.1m$, respectively. In python, these values form a two-dimensional grid with points at **grid_x, grid_y = [ norm.ppf( np.linspace( 0.05, 0.95, 10 ) ) ] * 2**. The probability distributions are obtained from the sigmoid activation function values for the output neurons and therefore correspond to the positive spin probability at each lattice site. The diagram in the upper left corner is calculated at $T = 2.0$ while the diagrams in the upper right, lower left, and lower right quadrants are obtained at $T = 2.375$, $T = 2.5$ and $T = 3.0$ respectively, ($T = 2.375$ approximates the temperature at which the $8 \times 8$ spin system exhibits a specific heat maximum).

The structure of the 100 spin distributions in each subfigure of Figure 3 becomes more visible if, as in Figure 1 a spin of $+1$ is assigned to each output location for which the output probability is larger than $0.5$ and a spin of $-1$ to the remaining lattice locations. This yields the four diagrams of Figure 4 together with the magnetization and energy distributions displayed in Figure 5 and Figure 6, respectively. The distributions generated by points that are adjacent in latent coordinates differ incrementally as can be seen by following either a row or a column of any of the four diagrams in the sets of figures. Figure 4 possesses the additional feature that below the critical temperature, the synthetic realizations are dominated by two regions of highly magnetized states separated by a narrow region of partially magnetized states. Near the transition temperature, the region of partially magnetized states widens and the spin distributions in the intermediate region interpolate between a large cluster of one spin orientation and a large cluster of the opposite orientation. Therefore, the CL VAE transforms a spin distribution into patterns with high frequency spatial components mediated by the small diameter clusters. The wide regions of intermediate cluster sizes at higher temperatures further indicates that at higher temperatures, a broad spectrum of spin patters are present.

The patterns generated by points in the latent value space are dependent on the internal structure of the VAE. For example, Figure 7 displays the probability distributions corresponding to those of Figure 3 but for the DL VAE. Since in the DL procedure the 64 inputs are first flattened into a single vector and the original $8 \times 8$ lattice only reconstructed in the last layer, the ordered pattern of the output of a CL VAE is replaced by a series of vertical stripes. However, Figure 8, which depicts the DL VAE energy distribution analogous to the CL result of Figure 6 demonstrates that the region of latent vector space associated with intermediate magnetizations again increases rapidly with increasing temperature close to and above the critical transition temperature, reproducing the qualitative behavior of the CL network.

**Conclusions and Future Directions:** Although the RBM, PCA and VAE generative algorithms have been successfully applied to image manipulation, this and previous papers have demonstrated that they do not accurately reproduce statistical quantities such as the specific heat in lattice theory applications unless the dimensionality and hence degrees of freedom of the intermediate representation (e.g. the number of hidden neurons, principle components or latent space variables) is comparable with the number of lattice sites. The lack of accuracy is especially apparent when contrasted with optimized Monte Carlo or renormalization group-based approaches.[68,72–91] On the other hand, even for small dimensional intermediate representations neural networks reproduce the qualitative behavior of the spin system. The latent representation of a VAE can further, as shown here, indicate the existence of clusters of different scales and the extent to which they contribute to a statistical system at a given temperature. While outside the scope of this paper, this result together with the known ability of the VAE to distinguish images of multiple categories indicates that in physical systems with multiple competing phases the VAE will assign each phase to a separate latent vector space region. The critical temperature for phase transitions and the distribution of cluster sizes for the various phases could then be estimated from the growth of the width of the boundary layer between the locations of the different phase regions with temperature.

An examination of the synthetic VAE realizations for 3 or more latent space variables could also be of further interest since the energy-magnetization distribution of the synthetic and training data was here found to be far more accurate than in the standard two latent variable case. In addition, the analysis systems with 3 or more coexisting phases could be significantly improved if the differing phase regions are more distinctly separated in higher dimensional latent spaces.

**Acknowledgements:** The Natural Sciences and Engineering Research Council of Canada (NSERC) is acknowledged for financial support.

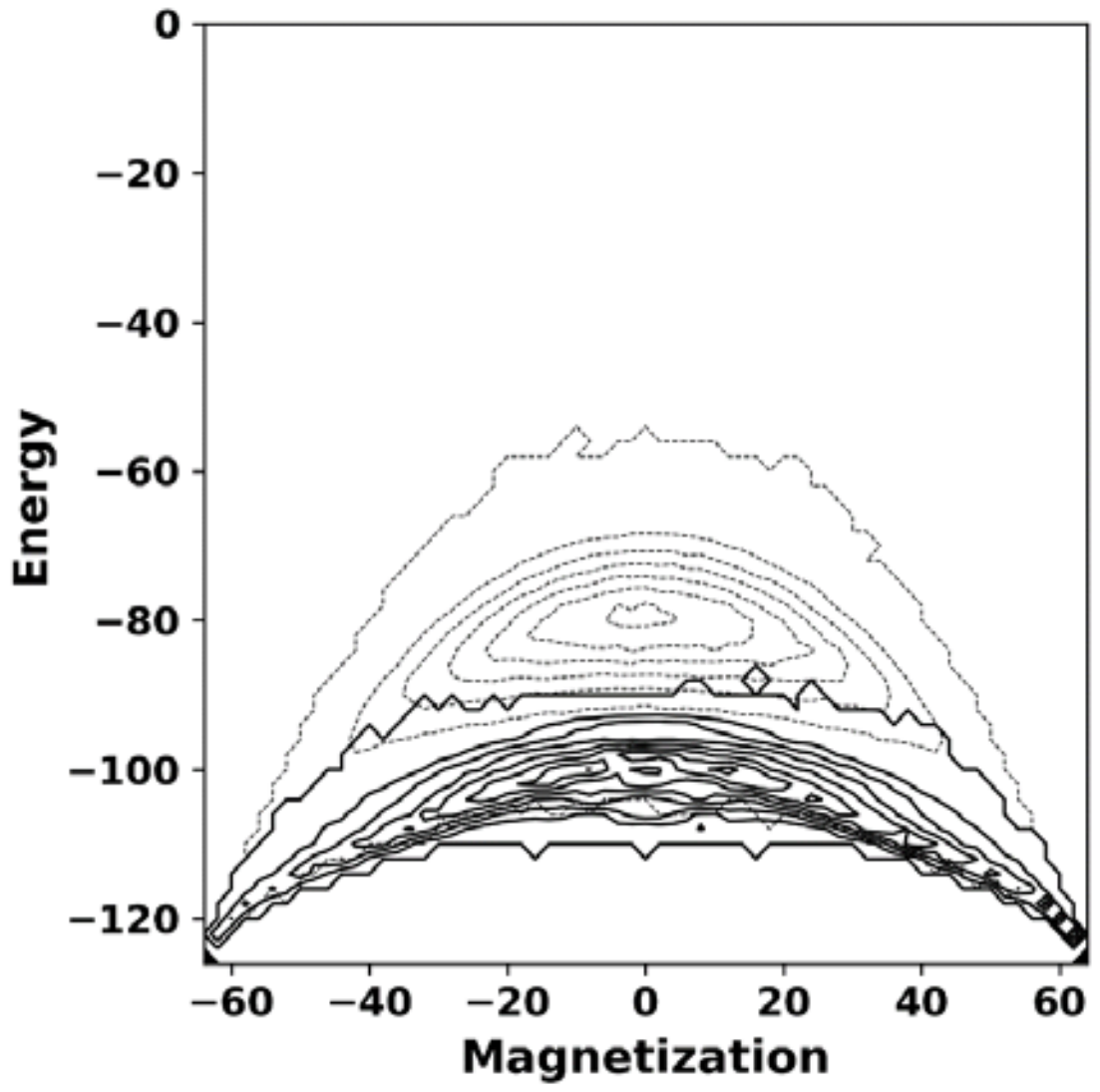

*Figure 1: The joint energy-magnetization distribution of the input $8 \times 8$ Ising model data for $T = 3.5$ (dashed lines) compared to the result of a VAE with 2 latent variables, where the VAE distribution is generated with the deterministic procedure described in the text.*

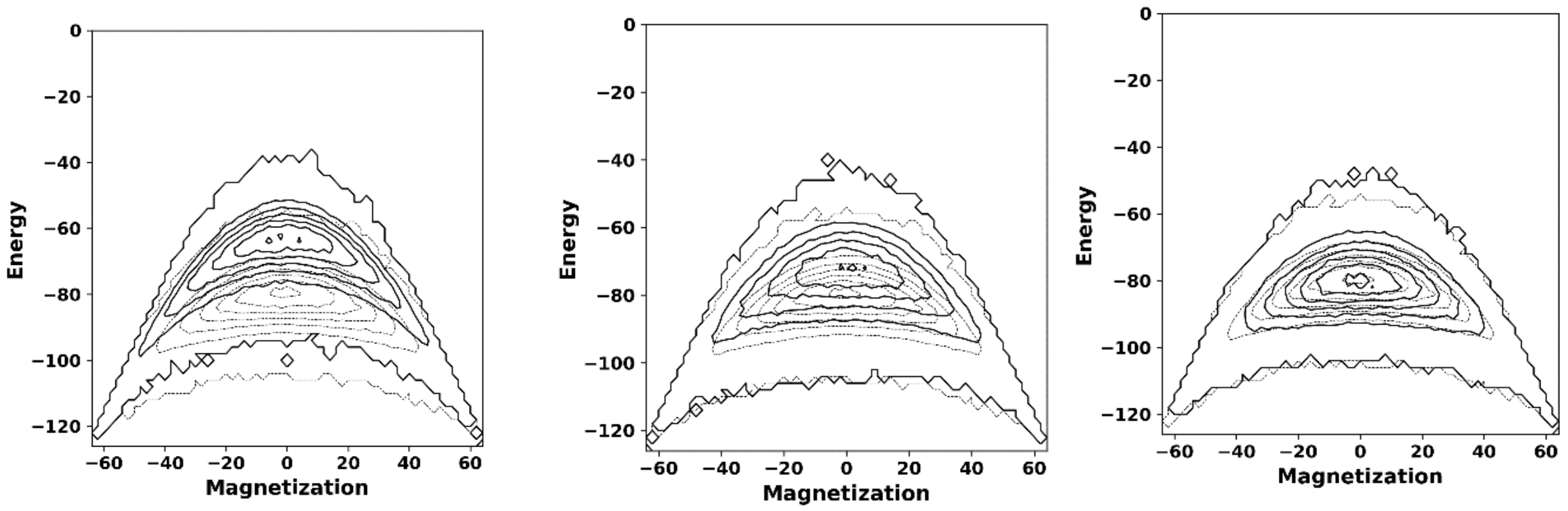

*Figure 2: The joint energy-magnetization distribution of the input $8 \times 8$ Ising model data at $T = 3.5$ (dashed lines) compared to the result of a VAE with 2, 3 and 10 latent variables (left, center and middle diagrams respectively), where the VAE results were generated with the probabilistic procedure described in the text.*

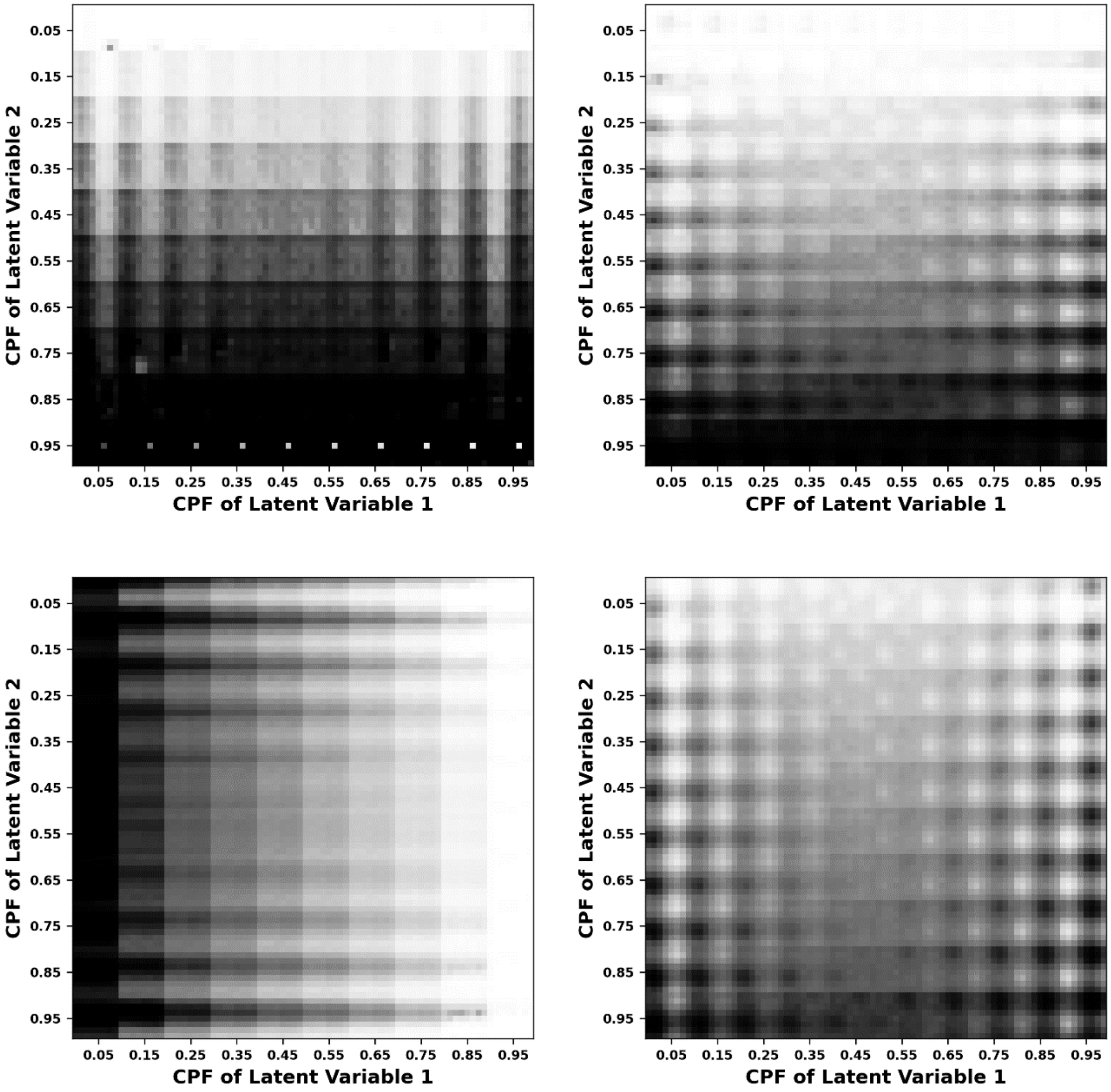

*Figure 3: The $8 \times 8$ synthetic convolutional layer (CL) VAE probability distributions for a $10 \times 10$ set of values in the latent vector space that are equally spaced in the cumulative probability function of the normal Gaussian distribution. The temperatures are $T = 2.0, 2.375, 2.5, 3.0$ for the upper right, upper left, lower left, and lower right quadrants, respectively.*

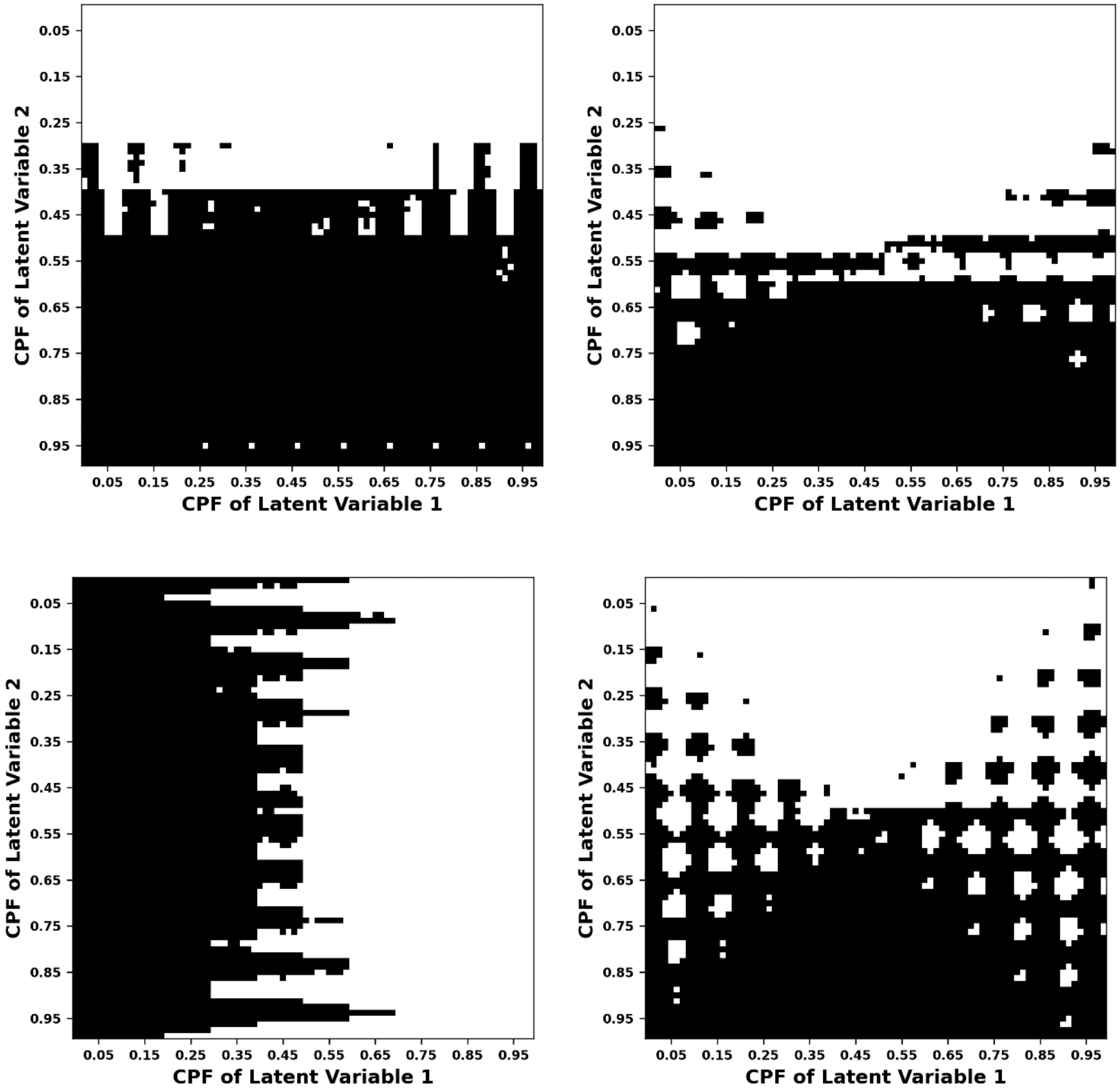

*Figure 4*: As in the previous figure, but with a spin of $+1$ *assigned to each output location for which the VAE output probability is larger than* $0.5$ *and a spin of* $-1$ *to the remaining lattice locations*.

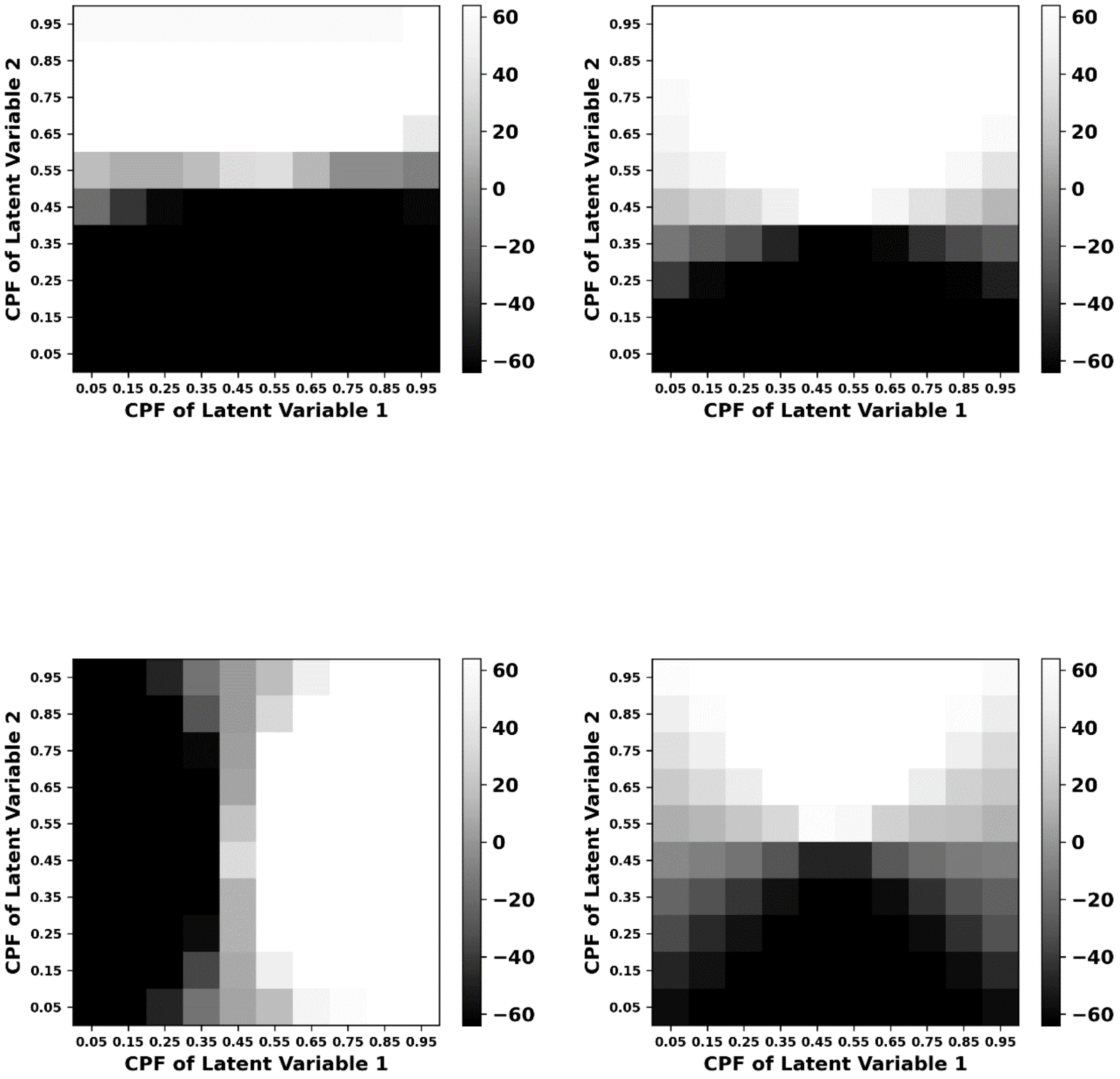

*Figure 5: The magnetization distributions associated with Figure 4*

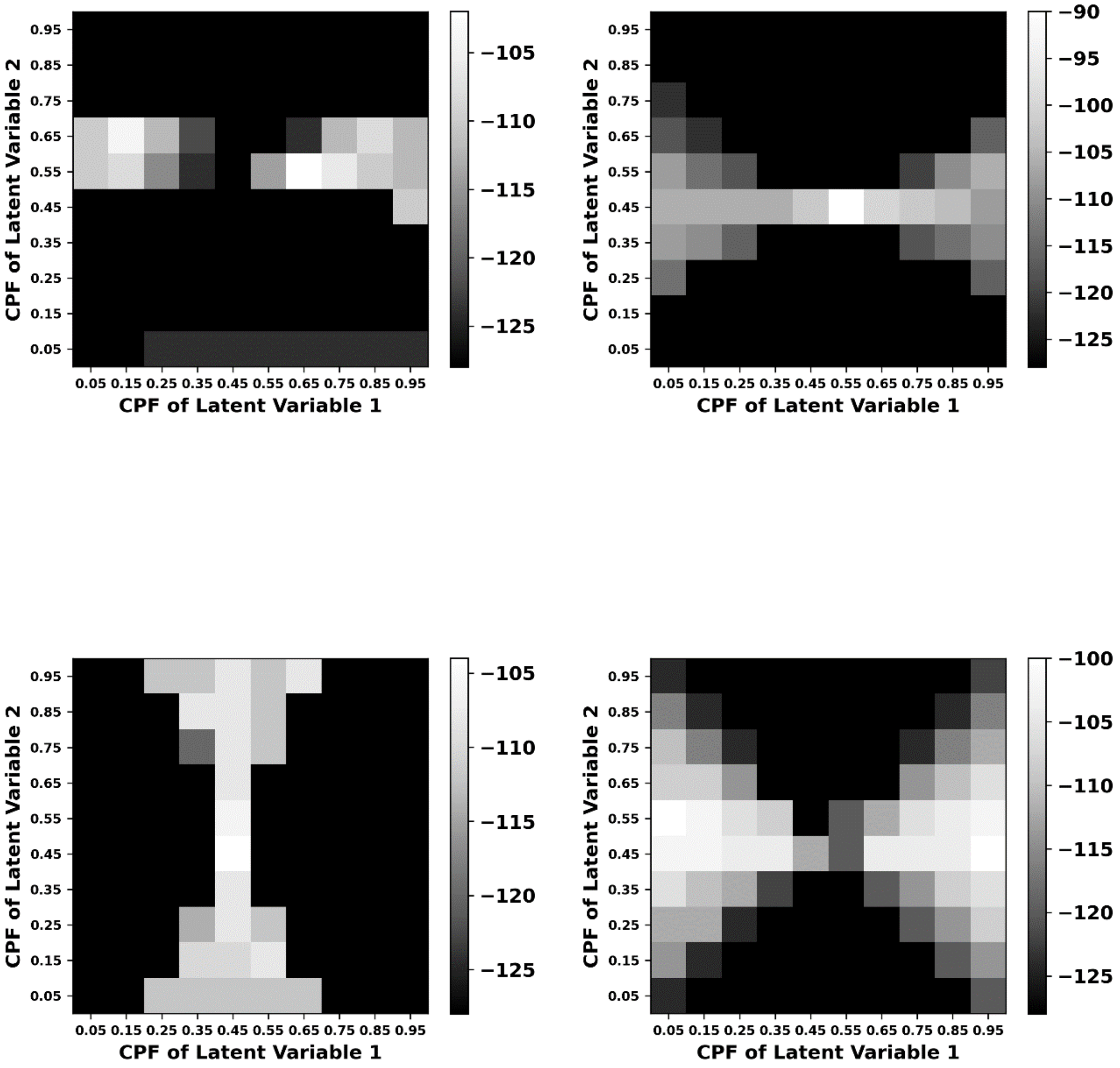

*Figure 6: The The energy distributions associated with Figure 4*

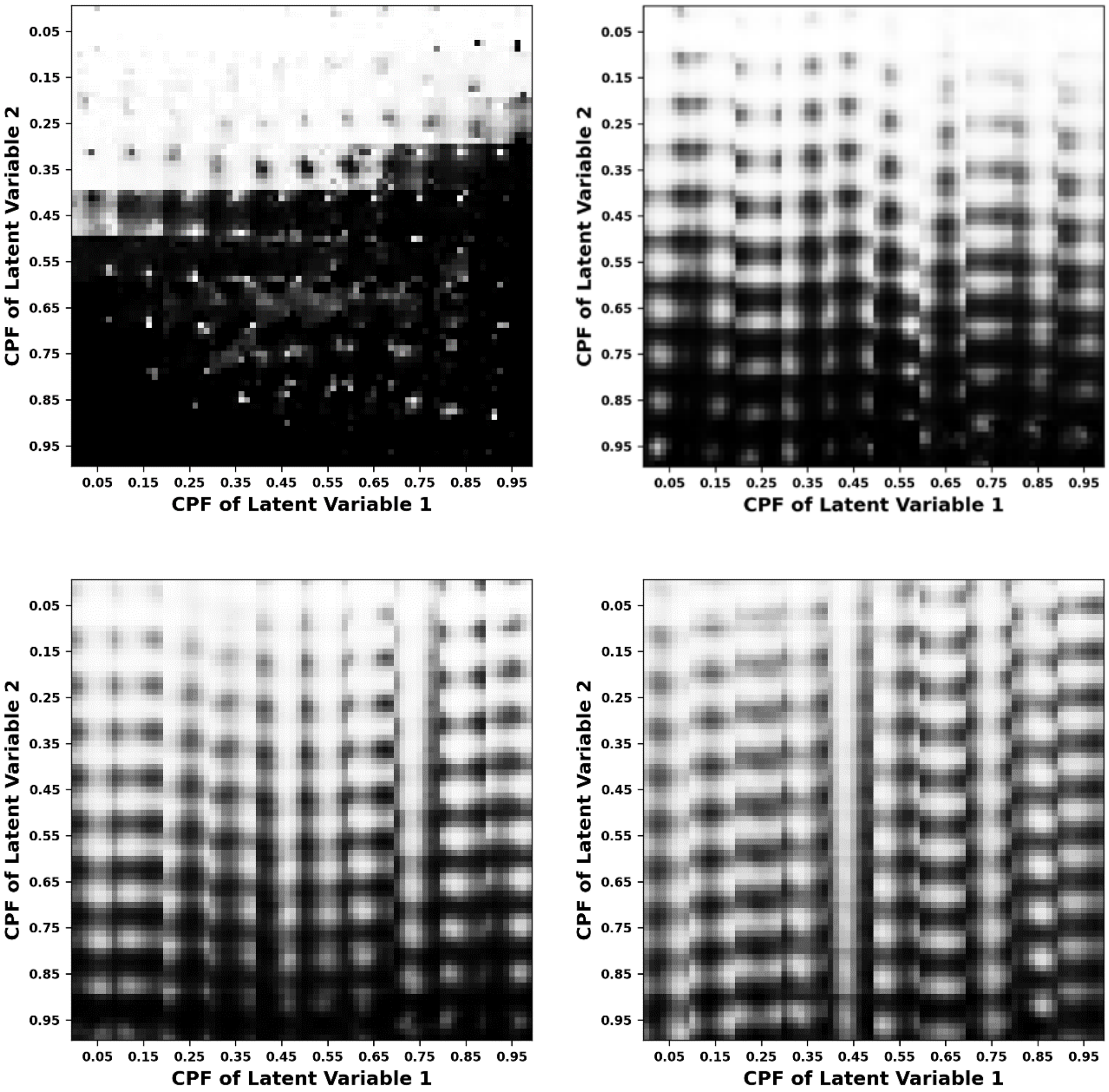

*Figure 7: The probability distributions corresponding to those of Figure 3 but for the DL implementation of the VAE.*

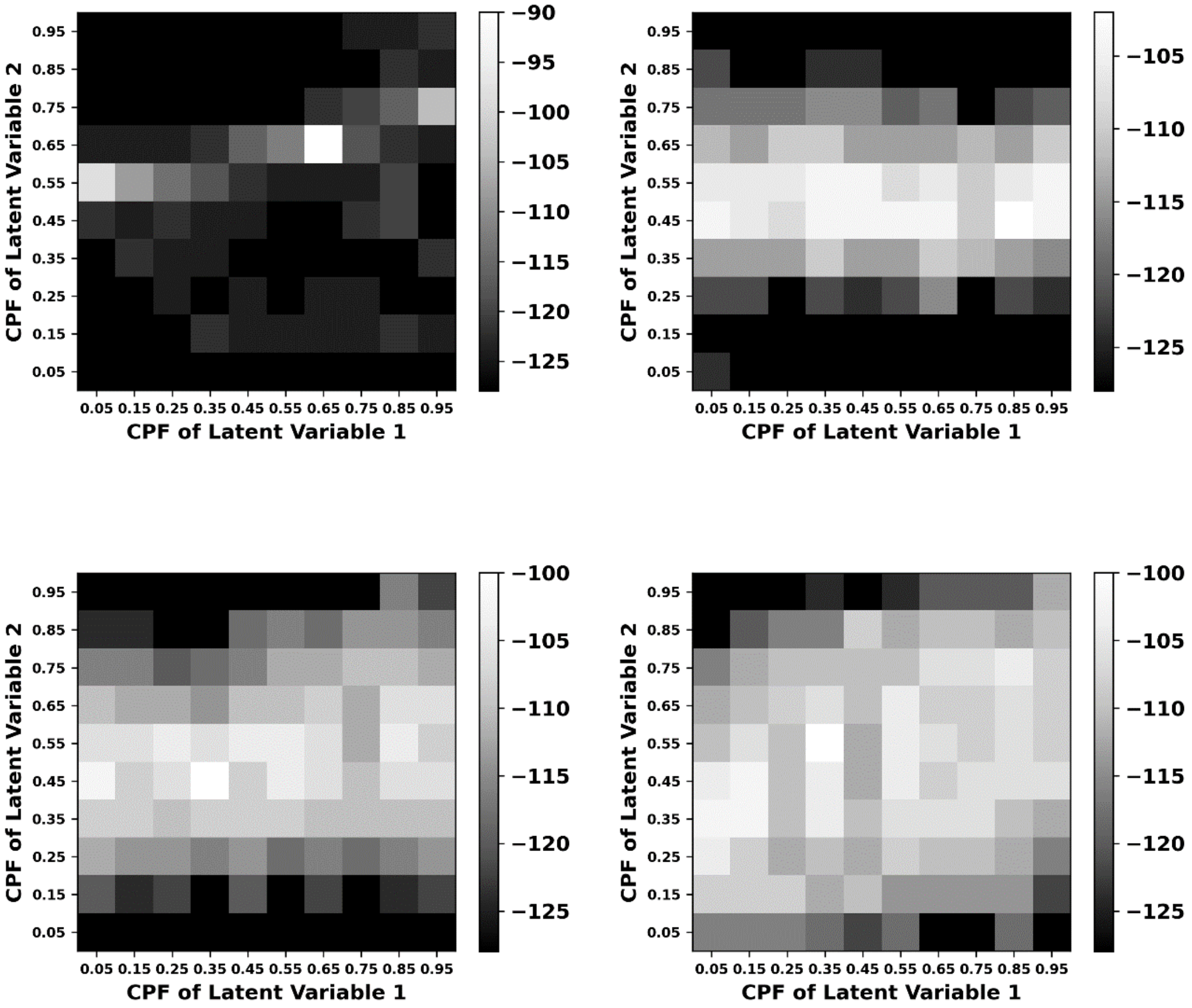

*Figure 8: The energy distributions corresponding to those of Figure 6 but for the DL implementation of the VAE.*